\documentclass[twocolumn,showkeys,pra]{revtex4}
\usepackage{epsf}
\usepackage{amsmath,amsthm,amssymb}
\usepackage{color}
\usepackage{dcolumn}
\usepackage{bm}
\usepackage{graphicx,epsfig}
\usepackage[dvipsnames]{xcolor}

\graphicspath{{fig/}}
\everymath{\displaystyle}
\begin{document} 

\title{Relativistic 
spin dynamics conditioned by
dark matter axions}

\author{A. J. Silenko}
\affiliation{Bogoliubov Laboratory of Theoretical Physics, Joint Institute for Nuclear Research,
Dubna 141980, Russia}
\affiliation{Institute of Modern Physics, Chinese Academy of
Sciences, Lanzhou 730000, China}
\affiliation{Research Institute for
Nuclear Problems, Belarusian State University, Minsk 220030, Belarus}

\keywords{axion; dark matter; relativistic quantum mechanics; pseudoscalar interaction; spin dynamics}

\begin{abstract}
The relativistic spin dynamics defined by the pseudoscalar field of dark matter axions is rigorously determined. The relativistic Hamiltonian in the Foldy-Wouthuysen representation is derived. It agrees with the previously obtained nonrelativistic Hamiltonians and the relativistic classical estimation of the axion wind effect. In the relativistic Hamiltonian, the biggest term describes the extraordinary (three orders of magnitude) enhancement of the axion wind effect in storage ring experiments as compared with experiments with immobile particles. This term defines the spin rotation about the longitudinal axis. The effects caused by the axion-induced oscillating EDM and the axion wind consist in the spin rotations about the different horizontal axes and phases of stimulating oscillations differ by $\pi/2$. Experiments in a search for dark matter axions has been discussed. The two experimental designs for axion search experiments in storage rings have been elaborated.
\end{abstract}

\maketitle

The axion is a hypothetical neutral pseudoscalar particle with a very low mass $m_a<10^{-2}$eV/$c^2$. Its existence has been postulated by Peccei and Quinn \cite{PecceiQuinn1,PecceiQuinn2} to resolve the strong \emph{CP} problem in quantum chromodynamics. If the axion exists, it can be a possible component of cold dark matter. The mass of dark matter axions is restricted by astrophysical observations \cite{amassgamma,amassobs} and cosmological arguments \cite{amasstheor}. An experimental search for axions being quanta of the pseudoscalar field can result in a discovery of fifth force. The successful search needs the creation of the corresponding relativistic theory of spin dynamics.

We use the standard denotations of Dirac matrices (see, e.g., Ref. \cite{BLP}) and the system of units $\hbar=1,~c=1$. We include $\hbar$ and $c$ explicitly when this inclusion
clarifies the problem.

\emph{CP}-noninvariant interactions caused by dark matter axions are time-dependent. Like photons, moving axions form a wave which pseudoscalar field reads
\begin{equation} a(\bm r,t)=a_0\cos{(E_at-\bm p_a\cdot\bm r+\phi_a)}. \label{axion}
\end{equation} Here $E_a=\sqrt{m_a^2+\bm p_a^2}$, $\bm p_a$, and $m_a$ are the energy, momentum, and mass of axions \cite{GrahamPRDgen2018}. The Earth motion through our galactic defines its velocity relative to dark matter, $V \sim10^{-3}c$. Therefore, $|\bm p_a|\approx m_a V$ \cite{GrahamPRDgen} and axions and axion-like particles have momenta of the order of $|\nabla a|\sim 10^{-3}\dot{a}c$.

Axions manifest themselves in interactions with photons, gluons, and fermions (first of all, nucleons). The corresponding contributions to the Lagrangian density are defined by \cite{PospelovAxion,Grahamannurev,GrahamPRDgen,RevModPhysaxion,AxionEDMneutronPXR,ICHEP2018Axion}
\begin{equation}
\begin{array}{c}
{\cal L}_{\gamma}=-\frac{g_{a\gamma\gamma}}{4}aF_{\mu\nu}\widetilde{F}^{\mu\nu}=g_{a\gamma\gamma}a\bm E\cdot\bm B,\\{\cal L}_{g}=\frac{g_{QCD}^2C_g}{32\pi^2f_a}aG_{\mu\nu}\widetilde{G}^{\mu\nu},\qquad  {\cal L}_{N}=g_{aNN}
\gamma^\mu\gamma^5
\partial_\mu a,
\end{array}
\label{eq33nnw}
\end{equation} where $g_{QCD}^2/(4\pi)\sim1$ is the coupling constant for the color field, $C_g,\,g_{a\gamma\gamma}$ and $g_{aNN}$ are model-dependent constants, $f_a$ is the constant of interaction of axions with matter (axion decay constant), and the tilde denotes a dual tensor.

The approach based on Eq. (\ref{eq33nnw}) and following Eq. (\ref{aEDM}) is now generally accepted \cite{GrahamPRDgen2018,PospelovAxion,Grahamannurev,GrahamPRDgen,RevModPhysaxion,AxionEDMneutronPXR,ICHEP2018Axion,DiLuzioAxion,DzubaFPAxion,DereviankoAxion,StadnikFlambaumAxion,PhysRevLettCASPEr,BudkerRMP,AxionEDMPRD2019,FrozEDMPRD,Axion2020EPJC,StephensonAxion,KimSemertzidis}. However, we should note the existence of alternative approaches to the description of axions. In particular, we mention Refs. \cite{Nikitin,BalakinPopov,DvornikovSemikoz,GorbarV,SikivieAxion} defining axion-induced particle motion and spin evolution in electromagnetic and gravitational fields. The paper \cite{BalakinPopov} also considers the spin precession of relativistic charged particles in storage rings. While  one  follows  the  generally accepted approach, other approaches should also be taken into consideration under performing an experiment.

The constant $f_a$, the axion mass and their contribution to dark matter depend on time of a phase transition in the expanding Universe. This circumstance defines possibilities of a search for dark matter axions. If one supposes that all dark matter with the local density $\rho_{\rm DM}$ is formed by axions, the axion field amplitude (\ref{axion}) reads \cite{GrahamPRDgen}
\begin{equation}
a_0=\frac{\sqrt{2\rho_{\rm DM}}}{m_a}. \label{AxionAmplitude}
\end{equation}

The axion-photon interaction leads to the inverse Primakoff effect \cite{SikivieHaloscope,AnselmAxion}. This effect is observable and axions are searched in experiments with haloscopes. We can mention the experiments ADMX \cite{ADMX}, ADMX-HF \cite{ADMX-HF}, HAYSTAC \cite{HAYSTAC}, CAPP \cite{CAPPaxion,CAPP}, and RADES \cite{RADES} and the reviews \cite{Grahamannurev,RevModPhysaxion,ICHEP2018Axion,axionSemertzidisarxiv,PhysRevXCASPEr,DiLuzioAxion,SikivieAxion}. We can also note experiments with helioscopes (see, e.g., Ref. \cite{Helio}) and the light-shining-through-a-wall experiments \cite{LSW}.

Another possibility to search for axions is connected with their influence on the spin. Nucleon spins are effected, besides the direct interaction defined by the Lagrangian density ${\cal L}_{N}$, by the axion-gluon interaction. Its \emph{CP}-noninvariance results in an appearance of oscillating nucleon electric dipole moments (EDMs) which are proportional to the axion field. The corresponding Lagrangian density can be written as follows \cite{GrahamPRDgen,AxionEDMneutronPXR,Grahamannurev,KimSemertzidis}:
\begin{equation}\begin{array}{c}
{\cal L}_{aEDM}=-\frac i2 g_da
\sigma^{\mu\nu}\gamma^5
F_{\mu\nu},
\end{array} \label{aEDM}
\end{equation} where the EDM is equal to $d_a=g_da=g_da_0\cos{[m_a(t-\bm V\cdot\bm r)+\phi_a]}$.

We underline that Eq. (\ref{axion}) is used in this paper only for a specification of frequency of the axion field oscillation.

The two spin-dependent Lagrangian densities, ${\cal L}_{aEDM}$ and ${\cal L}_{N}$, substantially differ. The former Lagrangian density defines an interaction of oscillating EDM induced by the axion field with external electric and magnetic fields. The latter one depends only on the axion field and is independent of other fields. In the static limit, the corresponding Hamiltonians are given by (see Refs. \cite{GrahamPRDgen2018,RPJ}):
\begin{equation}\begin{array}{c}
{\cal H}_{aEDM}=-d_a\bm\sigma\cdot\bm E,\qquad {\cal H}_{N}=g_{aNN}(\bm\sigma\cdot\nabla a),
\end{array} \label{aEDMstl}
\end{equation} 
where $\bm\sigma$ is the Pauli matrix. The Hamiltonian ${\cal H}_{aEDM}$ defines the EDM effect. The Hamiltonian ${\cal H}_{N}$ is proportional to the momentum of axions (axion wind effect). If there is a constant magnetic field and the spin rotation frequency in this field is close to the axion field frequency $\omega_a=m_ac^2/\hbar$, the magnetic resonance leading to the spin depolarization takes place. The method of magnetic resonance has been used in the CASPEr experiment \cite{Grahamannurev,PhysRevXCASPEr,PhysRevLettCASPEr} and other spin experiments with immobile nucleons and nuclei \cite{AxionEDMneutronPXR,BudkerRMP}. Effects caused by both the oscillating EDM and the axion wind have been studied. In Ref. \cite{AntiprotonAxion}, a search for the interaction of
antiprotons with dark matter axions has been fulfilled in the Penning trap. Similar searches can be carried out for other antiparticles, such as positrons and antimuons \cite{AntiprotonAxion}. Other experiments searching for the axion wind have been analyzed in Ref. \cite{GrahamPRDgen2018}.

It has been obtained in Refs. \cite{PospelovAxion,StadnikFlambaumAxion} by the elimination method (developed by Pauli \cite{Pa}) that the \emph{nonrelativistic} Hamiltonian ${\cal H}_{N}$ contains also the term proportional to the \emph{particle} velocity $\bm v$:
\begin{equation}\begin{array}{c}
{\cal H}^{(1)}_{N}=g_{aNN}\dot{a}(\bm\sigma\cdot\bm v).
\end{array} \label{aEDMprl} \end{equation}
We underline that taking into account the term (\ref{aEDMprl}) in the experiment \cite{AntiprotonAxion} can be necessary despite of nonrelativistic velocities of antiprotons in this experiment. 

The new approach to the experimental search for dark matter axions based on storage ring experiments has been proposed relatively recently \cite{AxionEDMPRD2019}. Since particles in such experiments are relativistic, this approach needs a construction of relativistic quantum-mechanical dynamics of spin.
A purely classical consideration shows that the axion wind effect depends on an axion motion relative to a spinning particle \cite{FrozEDMPRD}. In the rest frame of a relativistic particle, the Hamiltonian ${\cal H}_{N}$ is proportional to the factor $\nabla a\sim\gamma m_a\bm v_a^{(0)}a$, where $\bm v_a^{(0)}$ is the axion velocity in this frame. After a transition to the laboratory frame, the $\gamma$ factor however cancels due to the time transformation \cite{FrozEDMPRD}. Thus, the axion wind effect amplifies $v_a^{(0)}/V\approx v/V\sim c/V\sim10^{3}$ times in storage ring experiments. This extraordinary enhancement is the main preference of the use of storage ring experiments for an axion search. However, any relativistic quantum-mechanical analysis has not been made in Ref. \cite{FrozEDMPRD} and other papers.
Some experimental methods of the axion search in storage rings have been elaborated in Refs. \cite{FrozEDMPRD,Axion2020EPJC,AxionEDMPRD2019,StephensonAxion,KimSemertzidis}.

Needed derivations are fulfilled in the present
study. The Lagrangian $L=\overline{\psi}{\cal L}\psi$ describing electromagnetic and pseudoscalar interactions of a Dirac particle is defined by
\begin{equation}\begin{array}{c}
{\cal L}=\gamma^\mu(i\hbar\partial_\mu-eA_\mu)-m+\frac{\mu'}{2}\sigma^{\mu\nu}F_{\mu\nu}-i\frac{d}{2}\sigma^{\mu\nu}\gamma^5F_{\mu\nu}\\
+g_{aNN}\gamma^\mu\gamma^5\Lambda_\mu,\qquad \Lambda_\mu=(\Lambda_0,\bm\Lambda)=\partial_\mu a=(\dot{a},\nabla a),\\ \gamma^5=\left(\begin{array}{cc} 0  &  -1 \\ -1 & 0 \end{array}\right),
\end{array} \label{electric}
\end{equation}  where $\mu'$ and $d=d_0+g_da$ are the anomalous magnetic and electric dipole
moments and $d_0$ is the constant EDM.

The corresponding Hamiltonian in the Dirac representation is given by
\begin{equation}\begin{array}{c}
{\cal H}=\beta m+\bm\alpha\cdot(\bm p-e\bm A)+e\Phi +\mu'(i\bm\gamma\cdot\bm E-\bm \Pi\cdot\bm B)\\-d(\bm \Pi\cdot\bm E+i\gamma\cdot\bm B)+g_{aNN}(-\gamma^5\Lambda_0+\bm\Sigma\cdot\bm\Lambda).
\end{array} \label{eqelect}
\end{equation}
This equation agrees with the nonrelativistic equation (\ref{aEDMstl}).

In the general case, it is convenient to present the Dirac Hamiltonian as follows:
\begin{equation}\begin{array}{c} {\cal H}=\beta{\cal M}+{\cal E}+{\cal
O},\qquad \beta{\cal M}={\cal M}\beta,\\ \beta{\cal E}={\cal E}\beta,
\qquad \beta{\cal O}=-{\cal O}\beta. \end{array} \label{eq3} \end{equation} The
even operators ${\cal M}$ and ${\cal E}$ and the odd operator
${\cal O}$ are diagonal and off-diagonal in two spinors,
respectively. Equation (\ref{eq3}) is applicable for a particle with any
spin if the number of components of a corresponding wave function
is equal to $2(2s+1)$, where $s$ is the spin quantum number.

To obtain the Schr\"{o}dinger form of relativistic quantum mechanics (QM), one needs to fulfill the relativistic Foldy-Wouthuysen (FW) transformation (see Refs. \cite{JMP,PRA,PRAnonstat,PRA2015,PRA2016} and references therein).
This is a relativistic extension of the transformation found by Foldy and Wouthuysen \cite{FW}.
The FW transformation is one of
basic methods of contemporary QM. An
importance of the FW transformation for physics has significantly
increased nowadays due to a great progress of the art of analytic
computer calculations. A great advantage of the FW representation
is a simple form of operators corresponding to classical
observables. In this representation, the Hamiltonian and all
operators are even, i.e., block-diagonal (diagonal in two
spinors). The passage to the classical limit usually reduces to a
replacement of the operators in quantum-mechanical Hamiltonians
and equations of motion with the corresponding classical
quantities. The possibility of such a replacement, explicitly or
implicitly used in practically all works devoted to the FW
transformation, has been rigorously proved for the stationary case
in Ref. \cite{JINRLett12}. The probabilistic interpretation of wave functions lost in the Dirac representation is restored in the FW one \cite{PRAFW}. Thanks to these properties, the FW
representation provides the best possibility of obtaining a
meaningful classical limit of relativistic QM not only for the
stationary case but also for the
nonstationary one \cite{PRAnonstat}.
The FW transformation is applicable not only for Dirac fermions but also for
particles with any other spins \cite{Case,PRDexact}.

Since our analysis is relativistic, we take into account all axion-dependent terms in Eq. (\ref{eqelect}). The corresponding terms in Eq. (\ref{eq3}) are given by
\begin{equation} \begin{array}{c}
{\cal M}=m-\mu'\bm \Sigma\cdot\bm B-d\bm \Sigma\cdot\bm E,\qquad {\cal E}=e\Phi+g_{aNN}\bm\Sigma\cdot\bm\Lambda,\\ {\cal O}=\bm\alpha\cdot(\bm p-e\bm A)+i\mu'\bm\gamma\cdot\bm E-id\bm\gamma\cdot\bm B-g_{aNN}\gamma^5\Lambda_0. \end{array} \label{eqMEO} \end{equation}

The relativistic FW-transformed Hamiltonian is defined by 
\cite{PRA,PRA2015}:
\begin{equation}\begin{array}{c}
{\cal H}_{FW}=\beta\epsilon+ {\cal E}+\frac 14\biggl\{\frac{1}
{2\epsilon^2+\{\epsilon,{\cal M}\}},\left(\beta\left[{\cal O},[{\cal O},{\cal
M}]\right]\right. \\ \left.-\left[{\cal O},[{\cal O},{\cal
F}]\right]\right)\biggr\},\quad {\cal F}={\cal E}-i\hbar\frac{\partial}{\partial
t},\quad\epsilon=\sqrt{{\cal M}^2+{\cal
O}^2}. \end{array} \label{eqfingn}
\end{equation}
It should be emphasized that Eq. (\ref{eqfingn}) is applicable for arbitrarily
strong external fields \cite{PRA,PRA2015}. However, we can restrict ourselves by the weak-field approximation.

The straightforward derivation results in the relativistic FW Hamiltonian
\begin{equation}\begin{array}{c} {\cal H}_{FW}={\cal H}_{1}+{\cal H}_{2}+{\cal H}_{3},\\
{\cal H}_{1}=\beta\epsilon'+e\Phi+\frac
   14\left\{\left(\frac{\mu_0m}{\epsilon'
   +m}+\mu'\right)\frac{1}{\epsilon'},\biggl(\bm\Sigma \cdot [\bm\pi
\times \bm E]\right.\\ \left.-\bm\Sigma \cdot [\bm E \times \bm\pi]\biggr)\right\} -\frac
12\left\{\left(\frac{\mu_0m}{\epsilon'}
+\mu'\right), \bm\Pi \cdot \bm B\right\} \\
+\frac{\mu'}{4}\left\{\frac{1}{\epsilon'(\epsilon'+m)},
\biggl[(\bm{B} \cdot \bm\pi)(\bm{\Pi} \cdot \bm\pi)+ (\bm{\Pi}
 \cdot \bm\pi)(\bm\pi \cdot \bm{B})
 \biggr]\right\} ,\\ {\cal H}_{2}=-d\bm\Pi \cdot \bm E+\frac{d}{4}\left\{\frac{1}{\epsilon'(\epsilon'+m)},
\biggl[(\bm{E} \cdot \bm\pi)(\bm{\Pi} \cdot \bm\pi)\right.\\ \left.+ (\bm{\Pi}
 \cdot \bm\pi)(\bm\pi \cdot \bm{E})\biggr]\right\}
-\frac d4\left\{\frac{1}{\epsilon'},\biggl(\bm\Sigma \cdot [\bm\pi
\times \bm B]\right.\\ \left.-\bm\Sigma \cdot [\bm B \times \bm\pi]\biggr)\right\},\qquad
\bm\pi=\bm p-e\bm A,
\end{array} \label{eq12} \end{equation}
where ${\cal H}_{1}$ defines the \emph{CP}-conserving part of the total Hamiltonian ${\cal H}_{FW}$, $\mu_0=e\hbar/(2m)$ is the Dirac magnetic moment, and $\epsilon'=\sqrt{m^2+\bm{\pi}^2}$.
We have omitted spin-independent terms defining contact interactions. The quantity ${\cal H}_{2}$ which has been first derived in Ref. \cite{RPJ} characterizes the contribution of the EDM. We should note that ${\cal H}_{2}$ contains additional terms as compared with Ref. \cite{RPJ} owing to the noncommutativity of $a$ with the operators $\nabla$ and $\partial/(\partial t)$. However, these terms are negligible due to a very small mass of the axion and are also omitted.

The new terms describing the interaction with the axion wind are given by
\begin{equation}\begin{array}{c}
{\cal H}_{3}=\frac{g_{aNN}}{2}\left\{\frac{\bm\Pi\cdot\bm p}{\epsilon'},\Lambda_0\right\}+\frac{g_{aNN}}{2}\biggl[\left\{\frac{m}{\epsilon'},\bm\Sigma\cdot\bm\Lambda\right\}\\ +\frac{(\bm\Sigma\cdot\bm p)}{\epsilon'(\epsilon'+m)}(\bm p\cdot\bm\Lambda)+(\bm\Lambda\cdot\bm p)\frac{(\bm\Sigma\cdot\bm p)}{\epsilon'(\epsilon'+m)}\biggr].
\end{array}
\label{eqaxion}
\end{equation} The first and biggest term in ${\cal H}_{3}$ obtained in the present study leads to an extraordinary increase of the axion wind effect in storage ring experiments by three orders of magnitude as compared with experiments with immobile particles. This term is essentially the relativistic generalization of earlier results obtained in Refs. \cite{PospelovAxion,StadnikFlambaumAxion}.

Equations (\ref{eq12}), (\ref{eqaxion}) confirm Eqs. (\ref{aEDMstl}), (\ref{aEDMprl}) previously obtained in the nonrelativistic approximation. Equation (\ref{eqaxion}) also approves the relativistic classical consideration of the axion wind effect presented in Ref. \cite{FrozEDMPRD}.

In the semiclassical approximation, the angular velocity of the spin rotation has the form
\begin{equation}\begin{array}{c} \bm\Omega_s=\bm\Omega_{1}+\bm\Omega_{2}+\bm\Omega_{3},\\ \bm\Omega_{1}=-\frac{e}{2m}
\left\{\left(g-2+\frac{2}{\gamma}\right)\bm B-
\frac{(g-2)\gamma}{\gamma+1}\bm\beta(\bm \beta\cdot \bm B)
\right. \\ \left.
-\left(g-2+\frac{2}{\gamma+1}\right)(\bm\beta\times\bm E)\right\},
\\
\bm\Omega_{2}=-\frac{e\eta}{2m}\left[\bm
E-\frac{\gamma}{\gamma+1}\bm\beta(\bm\beta\cdot\bm
E)+\bm\beta\times\bm
B\right],\\ \bm\Omega_{3}=2g_{aNN}\left[\Lambda_0\bm\beta+
\frac{\bm\Lambda}{\gamma}+\frac{\gamma}{\gamma+1}(\bm\beta\cdot\bm\Lambda)\bm\beta
\right]\\=-2g_{aNN}\sin{(m_at-\bm p_a\cdot\bm r+\phi_a)}a_0\left[m_a\bm\beta-
\frac{\bm p_a}{\gamma}\right.\\ \left.-\frac{\gamma}{\gamma+1}(\bm\beta\cdot\bm p_a)\bm\beta
\right],
\end{array}\label{eq15}\end{equation} where
$\bm\Omega_{1}$ is determined by the Thomas-Bargmann-Michel-Telegdi equation \cite{Thomas,BMT},  $\bm\beta=\bm v/c$, and the factors $g=4(\mu_0+\mu')m/e$ and $\eta=4(d_0+g_da)m/e$ are introduced.

The analysis of spin precession of a relativistic charged particle in storage rings has been carried out in Ref. \cite{BalakinPopov}. However, the result obtained disagrees with Eq. (\ref{eq15}). This is very natural because of the use of alternative approach in Ref. \cite{BalakinPopov}. As a rule, experimentalists apply the generally accepted approach.

Equation (\ref{eq15}) describes the spin motion relative to the Cartesian coordinate axes. In accelerators and storage rings, the spin dynamics is usually considered relative to the beam
direction. In this case, the angular velocity of spin rotation is given by \cite{FukuyamaSilenko} (see also Refs. \cite{RPJSTAB,JINRLettCylr} and references therein)
\begin{equation} \begin{array}{c}
\bm\Omega=\bm\Omega_s-\bm\Omega_c=-\frac{e}{m}\left[G\bm B-\left(G-\frac{1}{\gamma^2-1}\right)(\bm\beta\times\bm E)\right]\\
+\bm\Omega_{2}+\bm\Omega_{3}, \end{array} \label{eqcyc} \end{equation}
where $G = (g-2)/2$ and $\bm\Omega_c$ defines the angular velocity of the cyclotron motion. The (quasi)magnetic resonance takes place when $\Omega$ (more precisely, the vertical component of $\bm\Omega$) is close to the axion field frequency $\omega_a$. In this resonance, the interactions caused by the oscillating EDM and the axion wind effect act like a spin flipper. As follows from Eq. (\ref{eqcyc}), the resonance occurs under a definite connection between the axion mass and fields in storage rings. A comparison of oscillating terms in $\bm\Omega_{2}$ and $\bm\Omega_{3}$ shows that the relative importance of the axion wind effect is greater and less for a greater and a less axion mass, respectively.

It is now planned to search for dark matter axions in the frameworks of two EDM experiments which will be fulfilled in a magnetic storage ring with magnetic focusing \cite{EDMpaper} and in an electric storage ring with electric \cite{RevSciInstrum} or magnetic
\cite{HaciomerogluSemertzidis,HybridArXv} focusing. The use of a storage ring with main $E$ and $B$ fields has also been proposed \cite{AxionEDMPRD2019}. The JEDI collaboration will search for dark matter axions in a magnetic storage ring (see Refs.
\cite{AxionEDMPRD2019,Axion2020EPJC,StephensonAxion}). Varying the quantity $\Omega$ allows one to scan a studied interval of axion masses. In Refs.
\cite{AxionEDMPRD2019,Axion2020EPJC,StephensonAxion}, only the effect of oscillating EDM has
been considered because the large first term in Eq. (\ref{eqaxion})
was unknown. Other terms in this equation are comparatively
small and oscillate with the frequencies $\Omega\pm\Omega_c$.
In a magnetic storage ring, $(e/m)GB\approx\omega_a$ near the resonance.
We expect that the contributions of the oscillating
EDM and the axion wind effect into the spin rotation are
comparable under this condition. The axion wind effect
conditioned by the first term in Eq. (\ref{eqaxion}) can be even dominant. In the considered case, the quantities $g_dm$ and $g_{aNN}$ are comparable. In units used, $e=\sqrt\alpha\approx1/\sqrt{137}$. The constants $g_d$ and $g_{aNN}$ are not dimensionless.
Evidently, the effects caused by the oscillating
EDM and the axion wind consist in the spin rotations about the different axes (radial and azimuthal, respectively) and phases of stimulating oscillations differ by $\pi/2$. The theory developed in Ref. \cite{EurophysLett} is perfectly applicable for a quantitative description of the \emph{general} case of (quasi)magnetic resonance, when both axion-induced effects are taken into account. The more special case when the spin rotates about one axis has been considered  with respect to the EDM experiment in Refs. \cite{Silenko:2017a,JEDI:2017a}.

It is proposed in Refs. \cite{AxionEDMPRD2019,Axion2020EPJC,StephensonAxion} to use a simultanious decrease of the vertical magnetic field and the beam momentum for scanning studied intervals of spin rotation frequencies and axion masses. The ring radius should be conserved. Another possibility \cite{AxionEDMPRD2019} consists in keeping the vertical magnetic field and applying a radial electric field. Simultaneous variations of the latter field and the beam momentum with the conservation of the ring radius allow one to search for the axions in a studied interval of their masses. The use of a radiofrequency Wien filter at the frequency of the sidebands of the axion and spin rotation ($g-2$) frequency, $f_{WF}=(\omega_a\pm \Omega)/(2\pi)$, has been recently proposed in Ref. \cite{KimSemertzidis}. This proposition relates only to the search for an oscillating EDM because added oscillating electromagnetic fields are useless for a search for the axion wind effect. The product of two sinusoidal functions, $a(t)\bm B(t)$ or $a(t)\bm E(t)$, gives an oscillation at a needed angular frequency $\Omega$.

It should be noted that ultralight axions ($f_a=\omega_a/(2\pi)<10^{-4}$ Hz or $f_a<1/\tau_{coh}$, where $\tau_{coh}\sim10^{4}$ s is the spin coherence time) can be discovered by methods used for a search of static EDMs. One should only take into account that the axion wind effect, unlike the EDM one, leads to the spin turn about the longitudinal direction \cite{AxionEDMPRD2019,FrozEDMPRD}.

An advantage of the experiment in an electric storage ring is an opportunity to search for light and ultralight axions \cite{AxionEDMPRD2019}. The experiment is fulfilled with protons. The proton spin is frozen ($\Omega=0$) at the momentum $p=mc/\sqrt G$ [see Eq. (\ref{eqcyc})]. For light axions, freezing the spin allows one to search for the resonance $\Omega=\omega_a$ at rather small frequencies $1/\tau_{coh}<f_a<1$ Hz. Higher frequencies are also manageable. It has been previously proposed \cite{AxionEDMPRD2019} to use the (quasi)magnetic resonance. In this case, the radial electric field and the beam momentum can be varied and the ring radius should be conserved. Certainly, the above-mentioned radiofrequency Wien filter at the frequency $f_{WF}=(\omega_a\pm \Omega)/(2\pi)$ can also be applied.
In this case, the mass of searched axions is not limited.
We underline a high sensitivity of the axion search in an electric storage ring.

We have different propositions for conducting experiments on the axion search in storage rings.
We propose to use a Wien filter with constant (but changeable) $\bm B$ and $\bm E=-\bm v\times\bm B$ fields. A previously proposed change of a main (magnetic or electric) field and the beam momentum \emph{during one run} covering a studied interval of frequencies is not fully appropriate. Essential changes of main field and beam parameters can uncontrollably vary beam dynamics during the run. As a result, they distort the resonance curve and decrease the experimental sensitivity. On the contrary, Wien filters do not affect beam dynamics and influence only the spin motion. Of course, they can cause the well-known systematic error \cite{frozenEDM}. Indeed, an appearance of a vertical electric field in a ring with \emph{magnetic} focusing influences the magnetic moment and can lead to a false EDM effect \cite{frozenEDM}. However, this possible effect is always proportional to a vertical electric field in the Wien filter and cannot imitate a resonance on the resonance curve characterizing the run. Moreover, the effect can be significantly decreased \cite{HybridArXv,frozenEDM}.

We can add that fields in a Wien filter can be inverted ($\bm B\rightarrow-\bm B,~\bm E\rightarrow-\bm E$).

Our another proposition consists in an application of a radiofrequency cavity stimulating coherent synchrotron oscillations of the beam at the frequency $f_{rf}=2\pi|\omega_a\pm \Omega|$. After the modulation, the beam velocity reads
\begin{equation} \begin{array}{c}
v(t)=v_0+\Delta v_{rf}\cos{(\omega_{rf}t+\phi_{rf})}\\+\Delta v_{fr}\cos{(\omega_{fr}t+\phi_{fr})}, \end{array} \label{mvelo} \end{equation} where $\Delta v_{rf},~\omega_{rf}=2\pi f_{rf}$, and $\phi_{rf}$ are the amplitude, angular velocity, and phase of the modulated synchrotron velocity oscillations and $\Delta v_{fr},~\omega_{fr}$, and $\phi_{fr}$ are corresponding parameters of the free synchrotron velocity oscillations. The achievable amplitude of the velocity oscillations is $3.5\times10^{6}$ m/s \cite{OMS} and therefore $\Delta v_{rf}/v_0$ is greater than 0.01. This ratio is roughly two orders of magnitude greater than the corresponding ratio of the effective field amplitude in the radiofrequency Wien filter to the main field in the storage ring in the method \cite{KimSemertzidis}. The latter quantities should be averaged over the ring circumference. Therefore, the application of our proposition for an axion-induced EDM search in a magnetic storage ring should increase the sensitivity two orders of magnitude as compared with the method \cite{KimSemertzidis}. Furthermore, our proposition (unlike the method \cite{KimSemertzidis}) ensures excellent conditions for a search for the axion wind effect. In an electric storage ring, our proposition can also be successfully used for such a search but is useless for a search for the axion-induced EDM. 

Our second proposition is very similar to the resonance method developed by Y. Orlov \cite{YOr} (see also Ref. \cite{OMS}) for the EDM experiment. He proposed to stimulate coherent synchrotron oscillations at the angular frequency $\Omega$. However, his method has not been accepted by the (sr)EDM collaboration since a nonlinearity of beam dynamics in storage rings can result in unwanted spin resonance effects. Unlike this method, similar oscillations at the angular frequency $|\omega_a\pm \Omega|$ cannot lead to such effects. We should specify that the storage (observation) time should not be less than $2\pi/\omega_a$.

A shoice between different experimental propositions depends on a sensitivity and systematic errors and needs further investigations.

In summary, we have rigorously described the relativistic spin dynamics defined by the pseudoscalar field of dark matter axions. The relativistic Hamiltonian in the FW representation has been derived for the first time. It agrees with the known \emph{nonrelativistic} Hamiltonians \cite{GrahamPRDgen2018,PospelovAxion,StadnikFlambaumAxion} and the \emph{relativistic} classical estimation of the axion wind effect \cite{FrozEDMPRD}. The biggest term in this Hamiltonian extraordinarily (by three orders of magnitude) amplifies the axion wind effect
in storage rings as compared with experiments with immobile particles. It describes the spin rotation about the longitudinal axis with the angular frequency $\omega_a$ defined by the axion mass. The effects caused by the axion-induced oscillating EDM and the axion wind consist in the spin rotations about the different axes (radial and azimuthal, respectively) and phases of stimulating oscillations differ by $\pi/2$. The experimental search for dark matter axions has been discussed. The two experimental designs for axion search experiments in storage rings have been elaborated.

The author is grateful to N.N. Nikolaev for helpful discussions and comments and acknowledges the support by the National Natural Science
Foundation of China (grants No. 11975320 and No. 11805242) and by the Chinese Academy of Sciences President's International Fellowship Initiative (grant No. 2019VMA0019).


\begin{thebibliography}{}

\bibitem{PecceiQuinn1}
R. Peccei, H. R. Quinn, CP Conservation in the Presence of Pseudoparticles, Phys. Rev. Lett. \textbf{38}, 1440 (1977).

\bibitem{PecceiQuinn2}
R. Peccei, H. R. Quinn, Constraints imposed by CP conservation in the presence of pseudoparticles, Phys. Rev. D \textbf{16}, 1791 (1977).

\bibitem{amassgamma}
S. J. Lloyd, P. M. Chadwick, and A. M. Brown, Constraining the axion mass through gamma-ray observations of pulsars, Phys. Rev. D \textbf{100}, 063005 (2019).

\bibitem{amassobs}
J. H. Chang, R. Essig, and S. D. McDermott, Supernova 1987A constraints on Sub-GeV dark
sectors, millicharged particles, the QCD axion, and an axion-like particle. J. High Energy Phys.
\textbf{09}, 051 (2018).

\bibitem{amasstheor}
J. Preskill, M. B. Wise, and F. Wilczek, Cosmology of the invisible axion, Phys. Lett. B \textbf{120},
127 (1983); L. F. Abbott and P. Sikivie, A cosmological bound on the invisible axion, Phys. Lett. B \textbf{120},
133 (1983); M. Dine and W. Fischler, The not-so-harmless axion, Phys. Lett. B \textbf{120}, 137 (1983).

\bibitem{BLP}
V. B. Berestetskii, E. M. Lifshitz, and L. P. Pitayevskii,
{\em Quantum Electrodynamics}, 2nd ed. (Pergamon, Oxford, 1982).

\bibitem{GrahamPRDgen2018}
P. W. Graham, D. E. Kaplan, J. Mardon, S. Rajendran, W. A. Terrano,
L. Trahms, and T. Wilkason, Spin precession experiments for light axionic dark matter, Phys. Rev. D \textbf{97}, 055006 (2018).

\bibitem{PospelovAxion}
M. Pospelov, A. Ritz, and M. Voloshin, Bosonic super-WIMPs as keV-scale dark matter, Phys. Rev. D \textbf{78}, 115012 (2008).

\bibitem{GrahamPRDgen}
P. W. Graham and S. Rajendran,
New observables for direct detection of axion dark matter, Phys. Rev. D \textbf{88}, 035023 (2013).

\bibitem{Grahamannurev}
P. W. Graham, I. G. Irastorza, S. K. Lamoreaux, A. Lindner,
and K. A. van Bibber, Experimental Searches for the Axion and Axion-Like Particles, Annu. Rev. Nucl. Part. Sci. \textbf{65}, 485 (2015).

\bibitem{RevModPhysaxion}
J. E. Kim, G. Carosi, Axions and the strong \emph{CP} problem, Rev. Mod. Phys. \textbf{82}, 557 (2010).

\bibitem{AxionEDMneutronPXR}
C. Abel \emph{et al.}, Search for Axionlike Dark Matter through Nuclear Spin Precession
in Electric and Magnetic Fields, Phys. Rev. X \textbf{7}, 041034 (2017).

\bibitem{ICHEP2018Axion}
Y. K. Semertzidis, Axion dark matter searches, Proceedings of Science (ICHEP2018), 729
(2018).

\bibitem{DiLuzioAxion}
L. Di Luzio, M. Giannotti, E. Nardi, and L. Visinelli, The landscape of QCD axion models,
Phys. Rep. \textbf{870}, 1 (2020).

\bibitem{KimSemertzidis}
On Kim and Y. K. Semertzidis, New method of probing an oscillating EDM induced by axionlike dark matter using an rf Wien filter in storage rings, Phys. Rev. D \textbf{104}, 096006 (2021).

\bibitem{PhysRevLettCASPEr}
D. Aybas, J. Adam, E. Blumenthal, A. V. Gramolin, D. Johnson, A. Kleyheeg, S. Afach, J. W. Blanchard, G. P. Centers, A. Garcon, M. Engler, N. L. Figueroa, M. G. Sendra, A. Wickenbrock, M. Lawson, T. Wang, T. Wu, H. Luo, H. Mani, P. Mauskopf, P. W. Graham, S. Rajendran, D. F. J. Kimball, D. Budker, and A. O. Sushkov, Search for Axionlike Dark Matter Using Solid-State Nuclear Magnetic Resonance, Phys. Rev. Lett. \textbf{126}, 141802 (2021).

\bibitem{BudkerRMP}
M. S. Safronova, D. Budker, D. DeMille,
D. F. J. Kimball, A. Derevianko, and C. W.
Clark, Search for new physics with atoms and molecules, Rev. Mod. Phys. \textbf{90}, 025008 (2018).

\bibitem{DzubaFPAxion}
V. A. Dzuba, V. V. Flambaum, and M. Pospelov, Atomic ionization by keV-scale pseudoscalar dark-matter particles,
Phys. Rev. D \textbf{81}, 103520 (2010).

\bibitem{DereviankoAxion}
A. Derevianko, V. A. Dzuba, V. V. Flambaum, and M. Pospelov, Axio-electric effect,
Phys. Rev. D \textbf{82}, 065006 (2010).

\bibitem{StadnikFlambaumAxion}
Y. V. Stadnik and V. V. Flambaum, Axion-induced effects in atoms, molecules, and nuclei: Parity nonconservation,
anapole moments, electric dipole moments, and spin-gravity and spin-axion momentum couplings, Phys. Rev. D \textbf{89}, 043522 (2014).

\bibitem{AxionEDMPRD2019}
S. P. Chang, S. Hac{\i}\"{o}mero\u{g}lu, O. Kim, S. Lee, S. Park, and Y. K. Semertzidis, Axionlike dark matter search using the storage ring EDM method, Phys. Rev. D \textbf{99}, 083002 (2019).

\bibitem{FrozEDMPRD}
P. W. Graham, S. Hac{\i}\"{o}mero\u{g}lu, D. E. Kaplan, Z. Omarov, S.
Rajendran, and Y. K. Semertzidis, Storage Ring Probes of Dark Matter and Dark Energy, Phys. Rev. D \textbf{103}, 055010 (2021).

\bibitem{Axion2020EPJC}
J. Pretz, S. P. Chang, V. Hejny, S. Karanth, S. Park, Y. Semertzidis, E. Stephenson, H. Str\"{o}her, Statistical sensitivity estimates for oscillating electric dipole moment measurements in storage rings, Eur. Phys. J. C \textbf{80}, 107 (2020).

\bibitem{StephensonAxion}
E. J. Stephenson, A Search for Axion-like Particles with a Horizontally
Polarized Beam in a Storage Ring, Proceedings of Science (PSTP2019) 018.

\bibitem{Nikitin}
A. G. Nikitin and O. Kuriksha, Symmetries of field equations of axion electrodynamics, Phys. Rev. D \textbf{86}, 025010 (2012).

\bibitem{BalakinPopov}
A. B. Balakin and V. A. Popov, Spin-axion coupling, Phys. Rev. D \textbf{92}, 105025 (2015).

\bibitem{DvornikovSemikoz}
M. Dvornikov and V. B. Semikoz, Evolution of axions in the presence of primordial magnetic fields, Phys. Rev. D \textbf{102}, 123526 (2020).

\bibitem{GorbarV}
E. V. Gorbar, K. Schmitz, O. O. Sobol, and S. I. Vilchinskii, Gauge-field production during axion inflation in the gradient expansion formalism, Phys. Rev. D \textbf{104}, 123504 (2021).

\bibitem{SikivieAxion}
P. Sikivie, Invisible Axion Search Methods, Rev. Mod. Phys. \textbf{93}, 015004 (2021).

\bibitem{SikivieHaloscope}
P. Sikivie, Experimental Tests of the "Invisible" Axion,
Phys. Rev. Lett. \textbf{51}, 1415 (1983) [Erratum: Phys. Rev. Lett. \textbf{52}, 695 (1984)].

\bibitem{AnselmAxion}
A. A. Anselm, Experimental test for arion -- photon oscillations in a homogeneous constant magnetic field, Phys. Rev. D
\textbf{37}, 2001 (1988).

\bibitem{ADMX}
S. J. Asztalos, G. Carosi, C. Hagmann, D. Kinion, K. van Bibber, M. Hotz, L. J Rosenberg, G. Rybka, J. Hoskins, J. Hwang, P. Sikivie, D. B. Tanner, R. Bradley, and J. Clarke, SQUID-Based Microwave Cavity Search for Dark-Matter Axions, Phys. Rev. Lett. \textbf{104}, 041301 (2010); J. Hoskins, J. Hwang, C. Martin, P. Sikivie, N. S. Sullivan, D. B. Tanner, M. Hotz, L. J Rosenberg, G. Rybka, A. Wagner, S. J. Asztalos, G. Carosi, C. Hagmann, D. Kinion, K. van Bibber, R. Bradley, and J. Clarke, Search for nonvirialized axionic dark matter, Phys. Rev. D \textbf{84}, 121302 (2011).

\bibitem{ADMX-HF}
F. Mallet, M. A. Castellanos-Beltran, H. S. Ku, S. Glancy, E. Knill, K. D. Irwin, G. C. Hilton, L. R. Vale, and K. W. Lehnert, Quantum State Tomography of an Itinerant Squeezed Microwave Field, Phys. Rev. Lett. \textbf{106}, 220502 (2011); S. Lamoreaux, K. van Bibber, K. Lehnert, G. Carosi, Analysis of single-photon and linear amplifier detectors for microwave cavity dark matter axion searches,
Phys. Rev. D \textbf{88}, 035020 (2013); T. M. Shokair, J. Root, K. A. Van Bibber, B. Brubaker, Y. V. Gurevich, S. B. Cahn, S. K. Lamoreaux, M. A. Anil, K. W. Lehnert, B. K. Mitchell, A. Reed, G. Carosi, Future Directions in the Microwave Cavity Search for Dark Matter Axions, Int. J. Mod. Phys. A \textbf{29}, 1443004 (2014).

\bibitem{HAYSTAC}
K. M. Backes, D. A. Palken, S. Al Kenany, B. M. Brubaker, S. B. Cahn, A. Droster, G. C. Hilton, S. Ghosh, H. Jackson, S. K. Lamoreaux, A. F. Leder, K. W. Lehnert, S. M. Lewis, M. Malnou, R. H. Maruyama, N. M. Rapidis, M. Simanovskaia, S. Singh, D. H. Speller, I. Urdinaran, L. R. Vale, E. C. van Assendelft, K. van Bibber, and H. Wang, A quantum enhanced search for dark matter axions, Nature \textbf{590}, 238
(2021).

\bibitem{CAPP}
S. Lee, S. Ahn, J. Choi, B. R. Ko, and Y. K. Semertzidis, Axion Dark Matter Search around
6.7 $\mu$eV, Phys. Rev. Lett. \textbf{124}, 101802 (2020); J. Jeong, S. W. Youn, S. Bae, J. Kim, T. Seong, J. E. Kim, and Y. K. Semertzidis, Search for Invisible Axion Dark Matter with a Multiple-Cell Haloscope, Phys.
Rev. Lett. \textbf{125}, 221302 (2020).

\bibitem{CAPPaxion}
O. Kwon, D. Lee, W. Chung, D. Ahn, H. S. Byun, F. Caspers, H. Choi, J. Choi, Y. Chong, H. Jeong, J. Jeong, J. E. Kim, J. Kim, \c{C}. Kutlu, J. Lee, M. J. Lee, S. Lee, A. Matlashov, S. Oh, S. Park, S. Uchaikin, S. W. Youn, and Y. K. Semertzidis, First Results from Axion Haloscope at CAPP around 10.7 $\mu$eV, Phys. Rev. Lett. \textbf{126}, 191802 (2021).

\bibitem{RADES}
A. \'{A} Melc\'{o}n, S. A. Cuendis, C. Cogollos, A. Díaz-Morcillo, B. D\"{o}brich, J. D. Gallego, B. Gimeno, I. G. Irastorza, A. J. Lozano-Guerrero, C. Malbrunot, P. Navarro, C. P. Garay, J. Redondo, T. Vafeiadis and W. Wuenschet, Axion searches with microwave filters: the RADES project, J. Cosmol.
Astropart. Phys. \textbf{05}, 040 (2018).

\bibitem{axionSemertzidisarxiv}
Y. K. Semertzidis and Sung Woo Youn, Axion Dark Matter: How to detect it? ArXiv:2104.14831 [hep-ph].

\bibitem{PhysRevXCASPEr}
D. Budker, P. W. Graham, M. Ledbetter, S. Rajendran, and A. O. Sushkov, Proposal for a Cosmic Axion Spin Precession Experiment (CASPEr), Phys. Rev. X \textbf{4}, 021030 (2014).

\bibitem{Helio}
V. Anastassopoulos \emph{et al.} (CAST Collaboration), New CAST limit on the axion-photon interaction, Nature Phys.
13, 584 (2017); A. Abeln \emph{et al.} (The IAXO collaboration), Conceptual design of BabyIAXO, the intermediate stage towards the International Axion Observatory. J. High Energy Phys. \textbf{2021}, 137 (2021).

\bibitem{LSW}
R. Ballou, G. Deferne, M. Finger, Jr., M. Finger, L. Flekova, J. Hosek, S. Kunc, K. Macuchova, K. A. Meissner, P. Pugnat, M. Schott, A. Siemko, M. Slunecka, M. Sulc, C. Weinsheimer, and J. Zicha, New exclusion limits on scalar and pseudoscalar axionlike particles from light shining through a wall, Phys. Rev. D \textbf{92}, 092002 (2015).

\bibitem{RPJ}
A.\,J. Silenko, Quantum-mechanical description of the
electromagnetic interaction of relativistic particles with
electric and magnetic dipole moments, Russ. Phys. J. \textbf{48},
788 (2005). 

\bibitem{AntiprotonAxion}
C. Smorra, Y. V. Stadnik, P. E. Blessing, M. Bohman, M. J. Borchert, J. A. Devlin,
S. Erlewein, J. A. Harrington, T. Higuchi, A. Mooser, G. Schneider, M. Wiesinger,
E. Wursten, K. Blaum, Y. Matsuda, C. Ospelkaus, W. Quint, J. Walz, Y. Yamazaki,
D. Budker, and S. Ulmer, Direct limits on the interaction of
antiprotons with axion-like dark matter, Nature \textbf{575}, 310 (2019).

\bibitem{Pa}
W. Pauli, Die allgemeinen Prinzipien der Wellenmechanik / Handbuch der Physik,
Vol. 24, Pt. 1, ed. by H. Geiger, K. Seheel (J. Springer, Berlin, 1933).


\bibitem{JMP}
A.\,J. Silenko, Foldy-Wouthuysen transformation for
relativistic particles in external fields, J. Math. Phys. {\bf 44}, 2952 (2003).

\bibitem{PRA}
A.\,J. Silenko, Foldy-Wouthyusen transformation and semiclassical limit for relativistic particles
in strong external fields, Phys. Rev. A \textbf{77}, 012116 (2008).

\bibitem{PRAnonstat}
A. J. Silenko, Energy expectation values of a
particle in nonstationary fields, Phys. Rev. A \textbf{91}, 012111 (2015).

\bibitem{PRA2015}
A.\,J. Silenko, General method of the relativistic Foldy-Wouthuysen transformation
and proof of validity of the Foldy-Wouthuysen Hamiltonian, Phys. Rev. A \textbf{91}, 022103 (2015).

\bibitem{PRA2016}
A.\,J. Silenko, General properties of the Foldy-Wouthuysen transformation and applicability of the corrected original Foldy-Wouthuysen method,
Phys. Rev. A \textbf{93}, 022108 (2016).

\bibitem{FW}
 L.\,L. Foldy, S.\,A. Wouthuysen, On the Dirac Theory of Spin 1/2
Particles and Its Non-Relativistic Limit, Phys. Rev. \textbf{78}, 29 (1950).

\bibitem{JINRLett12}
A. J. Silenko, Classical limit of relativistic quantum mechanical equations in the Foldy-Wouthuysen representation,
Pis'ma Zh. Fiz. Elem. Chast. Atom. Yadra \textbf{10},
144 (2013) [Phys. Part. Nucl. Lett. \textbf{10}, 91 (2013)].

\bibitem{PRAFW}
Liping Zou, Pengming Zhang, and A. J. Silenko, Position and spin in relativistic quantum mechanics,
Phys. Rev. A \textbf{101}, 032117 (2020). 

\bibitem{Case} K.\,M. Case, Some Generalizations of the Foldy-Wouthuysen Transformation,
Phys. Rev. \textbf{95}, 1323 
(1954).

\bibitem{PRDexact}
A. J. Silenko, High precision description and new properties of a spin-1 particle in a magnetic field,
Phys. Rev. D \textbf{89}, 
121701(R) (2014).

\bibitem{Thomas}
L. H. Thomas, The motion of the spinning electron, Nature \textbf{117}, 514 (1926);
The kinematics of an electron with an axis,
Phil. Mag. \textbf{3}, 1 (1927).

\bibitem{BMT}
V. Bargmann, L. Michel, and V. L. Telegdi, Precession of the polarization
of particles moving in a homogeneous electromagnetic field, Phys. Rev. Lett. \textbf{2}, 435 (1959).

\bibitem{FukuyamaSilenko}
T. Fukuyama and A.J. Silenko, Derivation of Generalized
Thomas-Bargmann-Michel-Telegdi Equation for a Particle with Electric
Dipole Moment, Int. J. Mod. Phys. A \textbf{28}, 1350147 (2013).

\bibitem{RPJSTAB}
A. J. Silenko, Equation of spin motion in storage rings in the cylindrical coordinate system.
Phys. Rev. ST Accel. Beams \textbf{9}, 
034003 (2006).

\bibitem{JINRLettCylr}
A. J. Silenko, Comparison of spin dynamics in the cylindrical and Frenet-Serret coordinate systems,
Phys. Part. Nucl. Lett. \textbf{12}, 8 (2015).

\bibitem{EDMpaper}
D. Eversmann \emph{et al.} (JEDI Collaboration), New Method for a Continuous Determination of the Spin Tune in Storage Rings and Implications for Precision Experiments, Phys. Rev. Lett.
\textbf {115}, 094801 (2015);
G. Guidoboni \emph{et al.} (JEDI Collaboration), How to Reach a Thousand-Second in-Plane Polarization
Lifetime with 0.97-GeV/\emph{c} Deuterons in a Storage Ring, Phys. Rev. Lett. \textbf{117}, 054801 (2016);
N. Hempelmann \emph{et al.} (JEDI Collaboration), Phase Locking the Spin Precession in a Storage Ring, Phys. Rev. Lett. \textbf{119}, 014801 (2017).

\bibitem{RevSciInstrum}
V. Anastassopoulos \emph{et al.}, A storage ring experiment to detect a proton electric dipole moment, Rev. Sci. Instrum. \textbf{87}, 115116 (2016).

\bibitem{HaciomerogluSemertzidis}
S. Hac{\i}\"{o}mero\u{g}lu and Y. K. Semertzidis, Hybrid ring design in the storage-ring proton electric dipole moment experiment, Phys. Rev. Accel. Beams \textbf{22}, 034001 (2019).

\bibitem{HybridArXv}
Z. Omarov, H. Davoudiasl, S. Hac{\i}\"{o}mero\u{g}lu, V. Lebedev, W. M. Morse, Y. K. Semertzidis, A. J. Silenko, E. J. Stephenson, R. Suleiman, Comprehensive Symmetric-Hybrid ring design for pEDM experiment at below $10^{-29}e\cdot$ cm, arXiv:2007.10332 [physics.acc-ph].

\bibitem{EurophysLett}
A. J. Silenko, General description of spin motion in storage rings in the presence of oscillating horizontal fields, Europhys. Lett. \textbf{118}, 61003 (2017).

\bibitem{Silenko:2017a}
A. J. Silenko, General classical and quantum-mechanical description of magnetic resonance: an application to electric-dipole-moment experiments, Eur. Phys. J. C {\bf 77}, 341 (2017).

\bibitem{JEDI:2017a}
A. Saleev 
\emph{et al.} (JEDI Collaboration), Spin tune mapping as a novel tool to probe the spin dynamics in storage rings, Phys. Rev. ST Accel. Beams \textbf{20}, 072801 (2017).

\bibitem{frozenEDM} F. J. M. Farley, K. Jungmann, J. P. Miller, W. M. Morse, Y. F. Orlov, B. L. Roberts, Y. K. Semertzidis, A. Silenko,
and E. J. Stephenson, New Method of Measuring Electric Dipole Moments in Storage Rings, Phys. Rev. Lett. {\bf 93}, 052001 (2004).

\bibitem{OMS}
Y. F. Orlov, W. M. Morse, and Y. K. Semertzidis, Resonance Method of Electric-Dipole-Moment Measurements in Storage Rings, Phys. Rev. Lett.
{\bf 96}, 214802 (2006).

\bibitem{YOr}
Y. F. Orlov, The Resonance Method of EDM Measurements in Storage
Rings: Conceptual Design, EDM in Storage Rings Internal Note No. 69, 2004
(unpublished); Resonance method of EDM measurements in storage rings, STORI 2005 Conference Proceedings, Schriften des
Forschungszentrums J$\ddot{\rm u}$lich, Matter and Materials, Vol.
30 (2005), pp. 223--226.


\end{thebibliography}
\end{document}